# Nondestructive real-space imaging of current density distributions in randomly networked conductive nanomaterials

*Takahiro Morimoto\*, Seisuke Ata, Takeo Yamada and Toshiya Okazaki\**

CNT-Application Research Center, National Institute of Advanced Industrial Science and Technology (AIST), Tsukuba 305-8565, Japan

Corresponding Authors: t-morimoto@aist.go.jp, toshi.okazaki@aist.go.jp

**ABSTRACT**

For realization the new functional materials and devices by conductive nanomaterials, how to control and realize the optimum networks structures are import point for fundamental, applied and industrial science. In this manuscript, the nondestructive real-space imaging technique has been studied with the lock-in thermal scope via Joule heating caused by ac bias conditions. By this dynamical method, a few micrometer scale current distributions are visualized in a few tens of minutes due to the frequency-space separation and the strong temperature damping of conductive heat components. Moreover, in the tensile test, the sample broken points were completely corresponding to the intensity images of lock-in thermography. These results indicated that the lock-in thermography is a powerful tool for inspecting the intrinsic network structures, which are difficult to observe by conventional imaging methods.

**INTRODUCTION**

In the last few decades the progress of analysis and synthesis technologies has led to the discovery of truly various nanomaterials, such as carbon nanotubes (CNTs)[1], graphenes[2], semiconducting nanowires[3], transition metal dichalcogenides (TMDs)[4,5] and metal nanoparticles[6–8]. These materials have unique properties originating from their geometrical structures, confined electronic structures and large specific surface areas. They are therefore attracting much attention as candidate materials for making new devices and materials having properties superior to those of conventional devices and materials, such as high-mobility devices[9,10], flexible transparent devices[11,12] and materials with high thermal and chemical resistances[13]. In many cases the nanomaterials consist of components randomly dispersed and oriented in two- or three-dimensional network structures called percolation states[14]. For maximization of the various properties, how to control and analyze the geometric structures, surface conditions, junctions and space distributions are obviously important points if we are to realize the new functional devices and materials[15,16]. Especially in the case of electric and optical devices, visualization of the real current path is quite useful for understanding the intrinsic mechanism of the devices and materials.

For visualization of network structures and current distributions, some methods based on scanning probe microscopy (SPM) have been already reported[17–25], *e.g.*, scanning gate microscopy (SGM) as a moving gate technique[19] and conductive atomic force microscopy (C-AFM) as a moving electrode technique[18]. These SPM-based techniques have successfully visualized local current density and resistance distributions in nanometer-scale imaging. It is difficult, however, to image whole devices and amounts of materials large enough to determine the bulk properties because of the long time needed for image accumulation. Moreover, these methods are applicable only to two-dimensional devices or thin materials because their mechanisms are sensitive to surface conditions and sample thickness. Therefore, a faster and noncontact method is needed for more efficient investigation of bulk properties originating from nanomaterial network structures.

Here we propose a new method for multiscale systems that is based on the real-time lock-in thermography (LIT) technique[26] with the Joule heating resulting from an ac bias conditions. The principal of this method was first suggested more than 40 years ago by Carlomagno[27], and in the 80s the camera based LIT technique was established in the nondestructive testing (NDT) of bulk materials[28,29]. For a long time, however, the limited resolution of infrared (IR) cameras restricted spread this useful method. In last decade the IR camera systems were dramatically improved, yielding more than a few hundred pixels in thermal microscope systems. Therefore, recently, in the field of failure analysis (FA) for integrated circuits, LIT is are gradually spreading as a fast testing method with a resolution higher than that of conventional laser scanning based methods[30]. The LIT technique is mainly used to improve sensitivity in FA and NDT because its signal-to-noise ratio (S/N) is higher than those of conventional plus or differential based methods[31]. In the nanomaterial research field, some manuscripts using a non-lock-in thermal microscope reported high-field breakdown in imaging with a heating condition[32,33]. In this study, however, we propose using this method to separate the wavelengths of various heat components originating from nanomaterial network structures in devices and materials (see details in the supporting information section S1). How to measure the heat component from only the nanomaterial networks is important for visualizing the current and resistance distributions in these randomly oriented systems.

**RESULTS AND DISCUSSION**

Figure 1a shows the schematic concept of the LIT method. If we apply a bias voltage to conductive samples, the heat power density ($P$) radiated should be given by a simple equation using the surface emissivity ($\varepsilon$), Stefan-Boltzmann constant ($\sigma$) and local temperature ($T$) ($P = \varepsilon \cdot \sigma \cdot T^4$). In this case, the heat power should be rapidly spread to and stored in base materials. Therefore, the time-dependent temperature components can be given by

$$T_{Total} = T_{Joule}(\theta) + T_{BG}(const) + T_{Diffuse}(\theta + \Delta) \qquad (1)$$

where $\theta$ is the phase of the ac bias voltage and $\Delta$ is the delay time caused by the thermal conduction. Hence, the first term represents the Joule heating having same frequency as the applied bias voltage, the second term is the almost constant heat component corresponding to the background heat storage, and the third term is the thermal diffusive component of the non-active region near the heat source. The information of nanostructures and current distributions might be contained in the first and third terms. In the case of dc bias voltage measurement, all heat component should be rapidly relaxed to constant background heat components, making it is impossible to measure the nanomaterial structures and local current density distributions. In the microscopic picture, the Joule heating originates from the electron-phonon interactions through the scattering phenomena. Therefore, the macroscopic Joule-heating signals are ensembles of scattering energy loss. For this reason, the intensity images obtained with the LIT method reflect the local power dissipation depending on the local voltage, resistance and current density in each position. The schematic images in Fig. 1b and c show the basic analysis process of the LIT method. By this method, the small heat components from the nanomaterials ($f = 0.1$– $100$ Hz) are separated from the background component ($f = $ constant) as shown in Fig. 1b. At this time, we use the ac bias frequency 25 Hz, which is maximum value for full-pixel-size measurement in our system. In the real-time LIT method, the measured thermal frame images are calculated under measurement and then the intensity images (R) and phase images ($\theta$) are reconstructed. These processes are similar to those in the conventional dual-phase lock-in measurement in optical or electrical measurements.

Figure 2a shows the optical microscope image of a tested sample. In this work we used as typical sample a conductive composite material that had relatively stable structures made of CNTs and polycarbonate (PC) base materials. Figure 2 b-e show IR and thermal images of the same sample shown in Fig. 2a. In infrared images obtained in non-bias conditions, the IR intensity is almost equal to the surface emissivity corresponding to the surface concentration distribution of nanomaterials and base materials. Therefore, the relatively uniform surface seen in the IR image means that CNTs distributions are relatively uniform at the surface of this sample. Figure 2c shows the thermal image obtained with the conventional accumulation method under the same ac bias voltages. In this image the detailed structures are completely smeared out by the background heat storage to the base materials. On the other hand, the LIT phase and intensity images were quite different, as shown in Fig. 2d and e. Some high-resistance cross-section lines are clearly observed as the low-intensity signals indicated by same-color arrow-pairs in Fig. 2d. These regions may be corresponding to weakly linked network structures originating from the CNT dispersion or junctions. Figure 2 f-h show higher-magnification images of areas around the higher-resistance regions. In these images, the clear hot spot pair structures are observed near the high-resistance cross-section lines. In the central region of these hot spots, the quite narrow current path linking hot spots is also observed as shown in Fig. 2 i-k. These structures indicate that the hot spot pair structures originate from the voltage concentration near the bottleneck structures of CNT networks. These results clearly show that the LIT method based on Joule heating detection can be used to visualize the internal current path and resistance distributions in complicated network structures consisting of randomly connected nanomaterials. This LIT method has some advantages over the conventional method based on SPM systems. The required measurement time is quite short, ranging from a few tens of seconds to a few tens of minutes, and is typically less than ten minutes when there is enough heat radiation and DC power (~mW). Moreover, the images were obtained under completely nondestructive conditions without heating damage. This is obviously important for reproducibility and additional measurement with other methods at the same position and under the same conditions.

Although the LIT images are made using the sample surface radiations, they are the ensembles of the signals delivered from the different depth positions within the thermal diffusion length ($\Lambda$), which is simply described as $\Lambda = \sqrt{2\alpha/f}$, where $\alpha$ is the thermal diffusivity and $f$ is the ac bias frequency. In the case of 25Hz measurement, the CNT composite materials show a $\Lambda$ greater than 2.0 mm. This means that a thin sample should show clearer LIT images in higher-magnification imaging. Figures 3a and b show the IR and optical microscope images of a sample less than 1 μm thick that was made by using a cryo-microtome. In these images, it is difficult to estimate the current path in these structures. On the other hand, the LIT image clearly shows the distribution of current density in the complicated network structures as shown in Fig. 3c. Figure 3d show the LIT intensity at various positions as functions of total current for the whole device. All signal intensities show clear dependence proportional to $I^2$ and this indicates that the signal should be originated from the only Joule heating by the LIT process. Another important point is the non-active region observed at the top-right of Fig. 3d. Although the CNTs exist in these regions, the LIT intensity signal is insignificantly small. This means that these CNTs should be electrically isolated from the current path and should not contribute to bulk properties. How to reduce the occurrence of non-active regions and realize uniform distributions is important for making bulk materials and devices with better properties. Therefore, the LIT technique should be useful for optimizing the microscopic structure and the sample fabrication process.

In thin-sample imaging, the resolution is dramatically better than bulk-sample results. The resolution of the LIT image of an isolated CNT is near the diffraction limit (~2.0 µm) of the wavelengths detected (from 3.0 to 5.0 µm). This resolution improvement is simply explained by strong damping of thermal conduction in ac bias conditions. In the case of a "thermally thin" sample (one with a thickness significantly smaller than the thermal diffusion length), the heat source should be considered as a mirror heat source chained by multiple reflections at the interfaces within the conduction path. Therefore, the temperature conduction is described as cylindrical thermal conduction from a line-like heat source by using Kelvin functions[26].

$$T(r,t) = A\left(ker\left(\frac{r\sqrt{2}}{\Lambda}\right) + ikei\left(\frac{r\sqrt{2}}{\Lambda}\right)\right)e^{i\omega t} \qquad (2)$$

Here $\Lambda$ is the thermal diffusion length as mentioned above, ω is the angular frequency ($f_{lock-in} = \omega/2\pi$), and $ker$ and $kei$ are Kelvin functions of the second kind. The LIT intensity should therefore be proportional to $\sqrt{(ker)^2 + (kei)^2}$. In the LIT, the intensity image could be separated in each phase component (see in detail in section S2). If we plot the faster in-phase component, only the Joule heating current Fpath is imaged as shown in Fig. 3e. Actually, in this image, the signal of the broken CNT is disappeared, and this means that this CNT signal consisted only of the thermal conduction component from the heat source. Therefore, it is possible to verify the quantitative analysis of the thermal conduction in these regions. Figure 3f shows the comparison between the LIT intensity profile and theoretical curves estimated from bulk thermal parameters (see in detail in section S3). In the case of thermal conduction from a heat source to FKM, the strongly damped curve has good correlations to the theoretical curve. On the other hand, the longer-distance signals of the CNT bundle show slight differences from the fitting curve estimated from the thermal conductivities of CNTs[34–37]. The more strongly damped curve of CNTs might originated from the thermal diffusion to FKM. Although the real temperature value is subtracted in the LIT process, these results clearly indicate that the LIT method is also useful for estimation of thermal properties in microscopic structures.

In the LIT images, the heat intensity should be influenced by the local electric structures depending on the CNTs connectivity as shown in Fig. 2. Therefore, in our method the distributions of current densities can be quantitatively analyzed through the intensity distribution. In Fig. 4 a-d the density-dependent results of CNT/FKM composite materials are shown for the same sample size and same applied DC power conditions ($P_{DC} = V_{DC} \times I_{DC}$). The inhomogeneity of LIT images is gradually smeared out by formations of uniform network structures by increasing CNT concentrations. In the case of the 5.0% sample, the intensity fluctuation is quite a bit smaller than those of lower-density samples. Reflecting these uniform network structure and CNT density increment, the sample resistances are also dramatically decreased, and peak shapes have narrow and near normal distributions as shown in Fig. 4e and f. For more quantitative analysis, the standard deviation (SD) are estimated from the intensity distributions (see in detail section S4). Figure 3g shows the SDs as a function of normalized resistance in samples with each CNT concentration. Although these normalized resistances still contained contact resistance, the SD has clear dependence on the $\log(R)$ originated from percolation behaviors in these systems[15,16]. By using this quantitative analysis, the distributions of nanomaterial network structures are easily estimated and visualized by this LIT method.

If the LIT images are really reflecting the whole current distributions and network structures, these signals should be directly correlated with physical-mechanical properties like those evaluated in the tensile testing of bulk materials. Figure 5 shows the tensile test results and LIT intensity images obtained at the same time in the same positions. The test piece was formed dumbbell-shaped to reduce stress concentration at the edge of the sample as shown in Fig. 5a. Before tensile testing (Fig. 5 b,c), the sample had two higher-thermal-intensity regions as shown in Fig. 5c and d. These strongly power-dissipated points mean that the local structure has bottlenecks of the current paths. These high-heating points consisted of weakly linked network structures corresponding to the existence of weakly linked CNT junctions. Therefore, these heating spots should be related to the physical-mechanical properties of bulk materials.

Figure 5e and f showed the IR and LIT images measured at 40% strained conditions, respectively. Applying strain in the horizontal direction obviously extended the sample length in the strain direction. This tensile effect is also observed in the I-V characteristics as shown in Fig. 5h. The sample resistance clearly increased from 25.9 k$\Omega$ to 35.4 k$\Omega$ at 40% strain. In this step, the LIT images also indicated difference from the initial state on their intensity profiles. The intensities of hot spots in the initial state are more emphasized by increasing local resistance. Moreover, the top of sample center region already showed dramatically decreased LIT intensity caused by broken internal network structures of CNTs. This prediction is confirmed in the next step results at 45% strain. In this condition, the sample started to break at two different points as shown in Fig. 5g. Both of them correspond to the higher LIT intensity region in the initial state. This means that the LIT intensity distributions are clearly related to physical-mechanical properties through the power dissipation distributions reflecting network structures of conductive nanomaterials. These results show that the LIT method will enable measurements to be made by another method before destructive tensile testing.

In this manuscript, the basic concept and typical results of the LIT method were discussed based on its use in conductive composite materials. This method has also wide applicability for conductive materials and devices. Therefore, we also report the results of its use on the low-dimensional graphene devices in a following manuscript[38].

**CONCLUSION**

We have investigated the LIT method as a nondestructive real-space method for imaging the current density, mechanical properties and space distributions for randomly networked nanomaterials. The small Joule heating of the nanomaterials was detected in frequency-space-separation images made using the lock-in thermography technique. These lock-in thermal images were directly correlated with current distributions and physical-mechanical properties reflecting inhomogeneity of conductive nanomaterial network structures. This is quite important for various investigations of the fundamental mechanisms carrier dynamics inside nanomaterial devices and materials. Moreover, the image accumulation faster than that of conventional SPM-based methods is also important in applied physics and industrial science for multiscale measurements at scales from microscopic to macroscopic. The lock-in thermography technique should be important tool for realization the more useful functional devices and materials with various nanomaterials.

**METHODS**

**The LIT measurement**

The LIT measurement was performed by using a lock-in thermography system for failure analysis (Themos-1000, Hamamatsu). IR signals with wavelengths from 3.0 to 5.0 μm were detected by the system's InSb camera. The electrical transport characteristics under an ac bias were evaluated by using a B1500 semiconductor parameter analyzer (Keysight). The electrodes were made from silver conductive pastes, and electrical probing was done using the home-made probe systems. The ac bias voltages were applied as rectangular signals having a 50% duty ratio. All the measurements were done at 25 Hz, whereas the camera's maximum frequency was 100 Hz. Therefore, four IR images were stored in each cycle of the lock-in process.

**Sample fabrication**

In the work reported in this manuscript, we used two different CNTs and two different base materials for composite samples. The super-growth and HiPCo single-walled CNTs were kneaded with a polycarbonate (PC, Panlight L-1225Y, Teijin Limited) and a fluoro-rubber (FKM, Daiel G912, Daikin Industries, Ltd.). The SG/PC samples were used in the experiment of Fig. 1 and Fig. 2. The SG/FKM samples were used in the measurement of Fig. 3 and Fig. 4. In the tensile test shown in Fig. 5, the HiPCo/FKM sample had strain behaviors suitable for the LIT measurement. The SG/PC samples were provided by a biaxial continuous kneader (Explore, MC-15TS, Explore Instruments) and formed with injection molding (Explore, 12cc, Explore Instruments). The SG/FKM and HiPCo/FKM samples were fabricated by a drop casting method. Except in the concentration-dependence measurement, the CNT concentration was 1.0 wt% to the base materials (PC and FKM). The thin sample measured in Fig. 4 was made by using the RM2265 cryo-microtome (Reica) at liquid-nitrogen temperature.

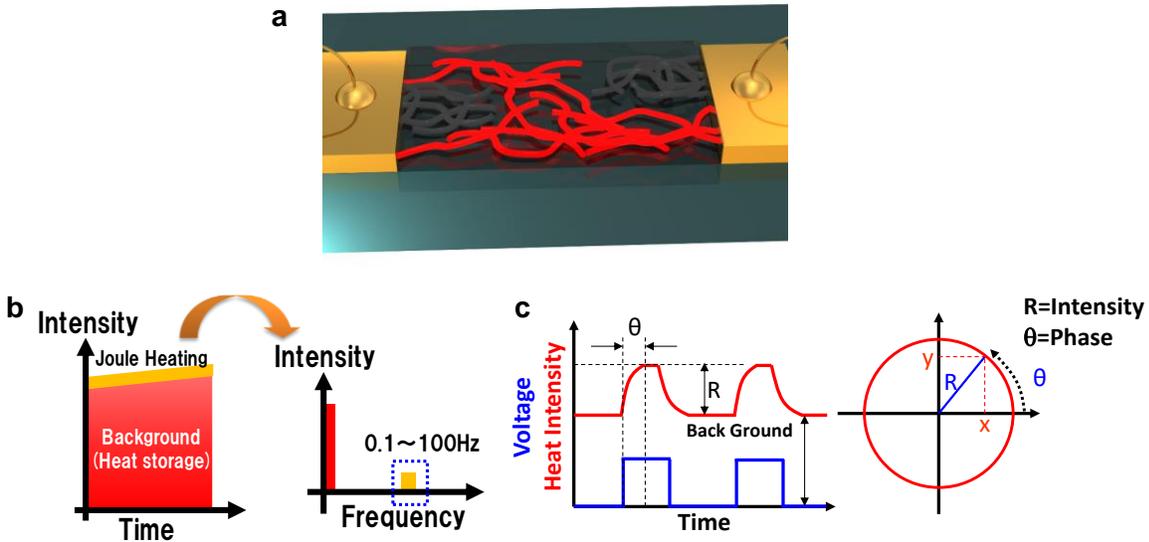

**Figure. 1.** (**a**) Schematic illustration of experimental concept of the LIT method. The heat generated in conductive nanomaterial network structures is radiated from the surface. (**b-c**) Schematic image of the voltage and heat sequence and the LIT signal processing. In the ac bias voltage condition, the Joule-heating component has a frequency different from that of background heat storage. The applied bias voltage generates Joule heat with some delay time at rise and drop points. After the lock-in process, the heat intensity and phase are represented as R and θ in a complex vector diagram.

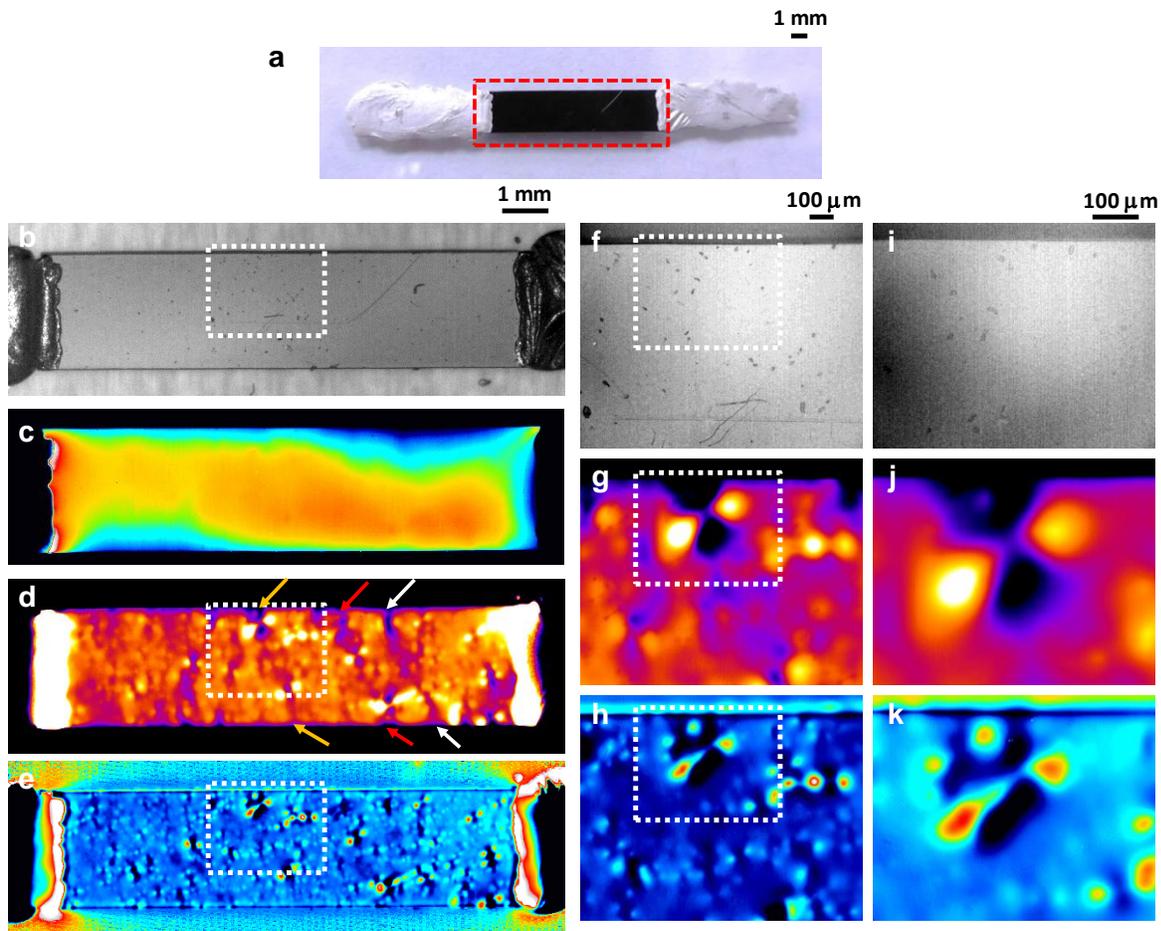

**Figure. 2** (**a**) Optical image of CNT/PC sample. Composite materials containing CNTs have an almost completely black surface, and the conventional optical method is not effective with these kinds of materials. (**b-e**) from top to bottom, low-magnification infrared, conventional-accumulation, intensity and phase images of the same sample shown in (a). In the infrared image, the sample showed a relatively flat and clean surface except for some particles and scratches. Moreover, the conventional method images showed only some blurred intensity distribution as shown in (c). The intensity (d) and phase (e) images show the heat distributions reflecting the network structures and local resistance. The pairs of arrow indicated in (d) indicate the higher-resistance cross sections across the current flow direction. (**f-k**) The middle- (f-h) and high- (i-k)

magnification images show the pair of hot spots near the high-resistance cross section shown in (d). The narrow conductive channel and voltage concentration are clearly observed in these results. All LIT images were obtained by applying 60-V and 6.4-mA bias conditions and using 10-minute accumulation.

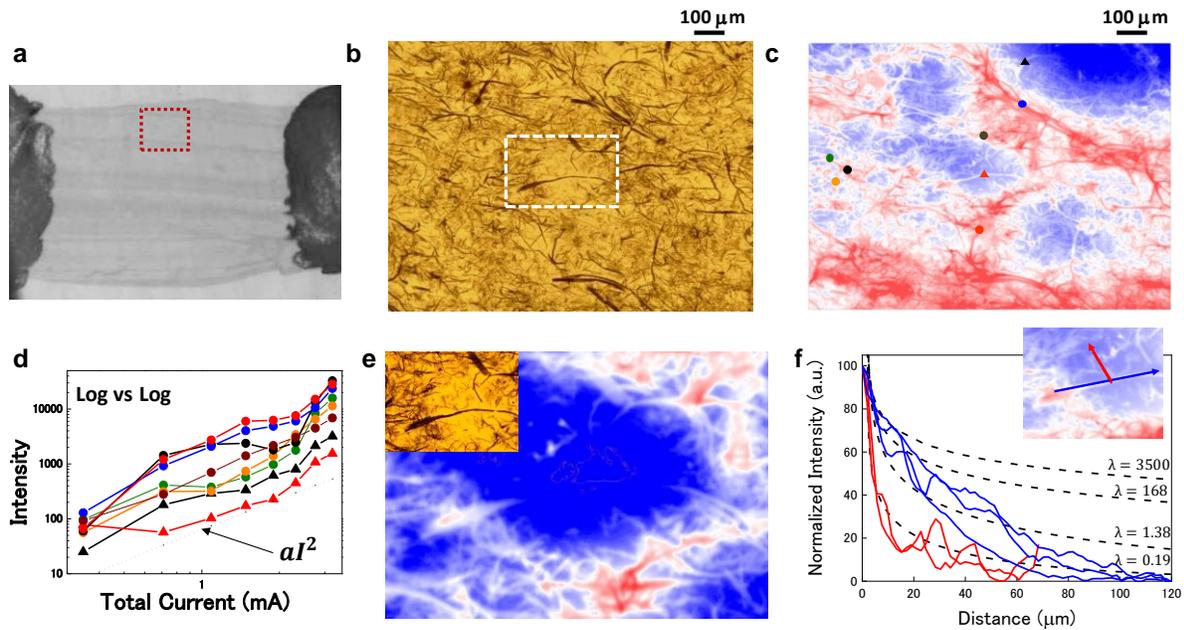

**Figure. 3** (**a**) Infrared image of the measured sample. The red-dotted square shows the measured area in (b,c). (**b**) Transmission optical microscope image at same area as c. The area in the white-dotted box is same area as shown in the inset of (e). (**c**) The LIT intensity image obtained with a high-magnification IR lens (x8, N.A.=0.75). The image area is 1200 μm x 960 μm. Each symbol corresponds to a measured point shown in (d). (**d**) The log-scale intensity as function of the log-scale total current. All the curves show the clear Joule-heating dependence of the total current. Here a is a pre-factor. (**e**) The phase-separation image for the in-phase component of ac bias voltage. The signal from the heat source disappeared in the heat-conductive region. The optical microscope image of the white-dotted area in (b) is shown in the inset. (**f**) The comparison between theoretical estimation curves and LIT intensity. The red curves represent the thermal conduction to the FKM. The blue curves represent the thermal conduction to CNTs. Dashed lines corresponding to the thermal conductivity of FKM (0.19), quartz (1.38), silicon (168) and a single

CNT (3500). These regions are chosen from the non-heat source regions as shown in (e). One example of chosen point at the FKM and CNT is shown in the inset.

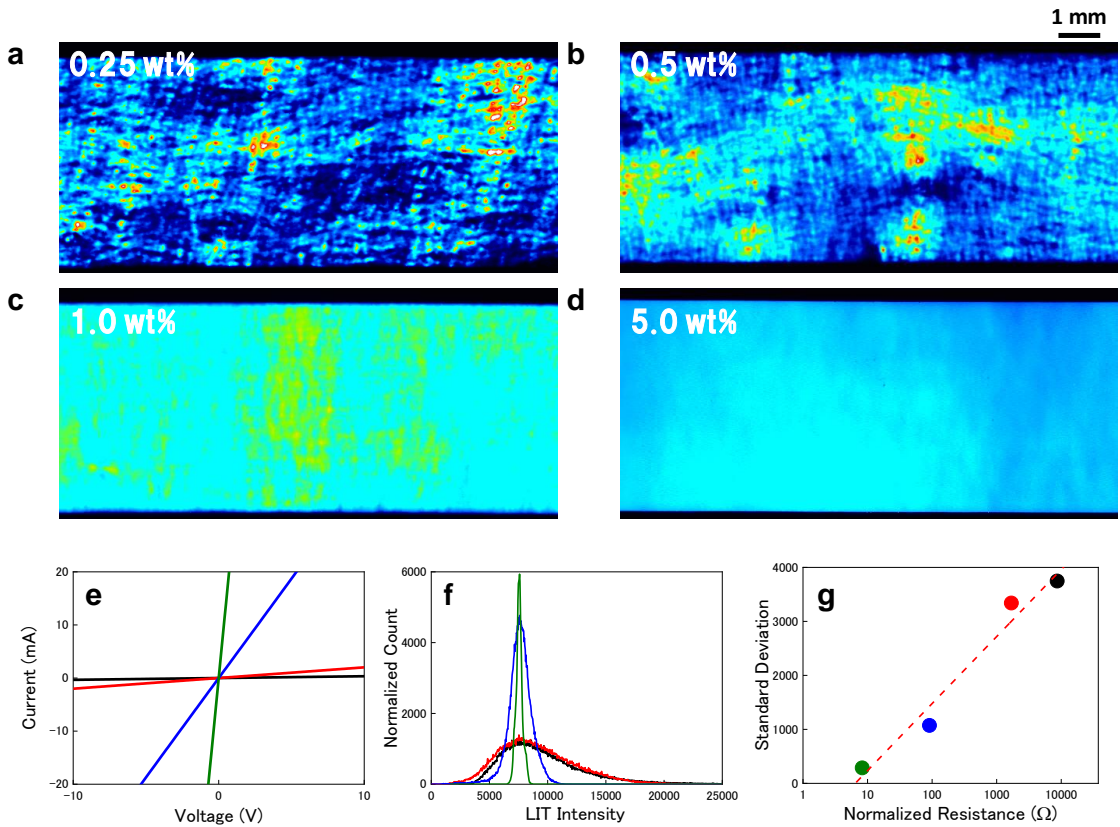

**Figure. 4 (a-d)** Density-dependent LIT intensity images for CNT contents from 0.25 wt% to 5.0 wt%. (**e**) The I-V characteristics of each sample. All data show clear linear dependence within the range of applied voltages. (**f**) Normalized count numbers shown as a function of the LIT intensity. The count numbers are normalized by the difference of sample size and image pixels. (**g**) Standard deviations of intensity distributions shown as a function of normalized sample resistance. The resistances are corrected by sample dimensions.

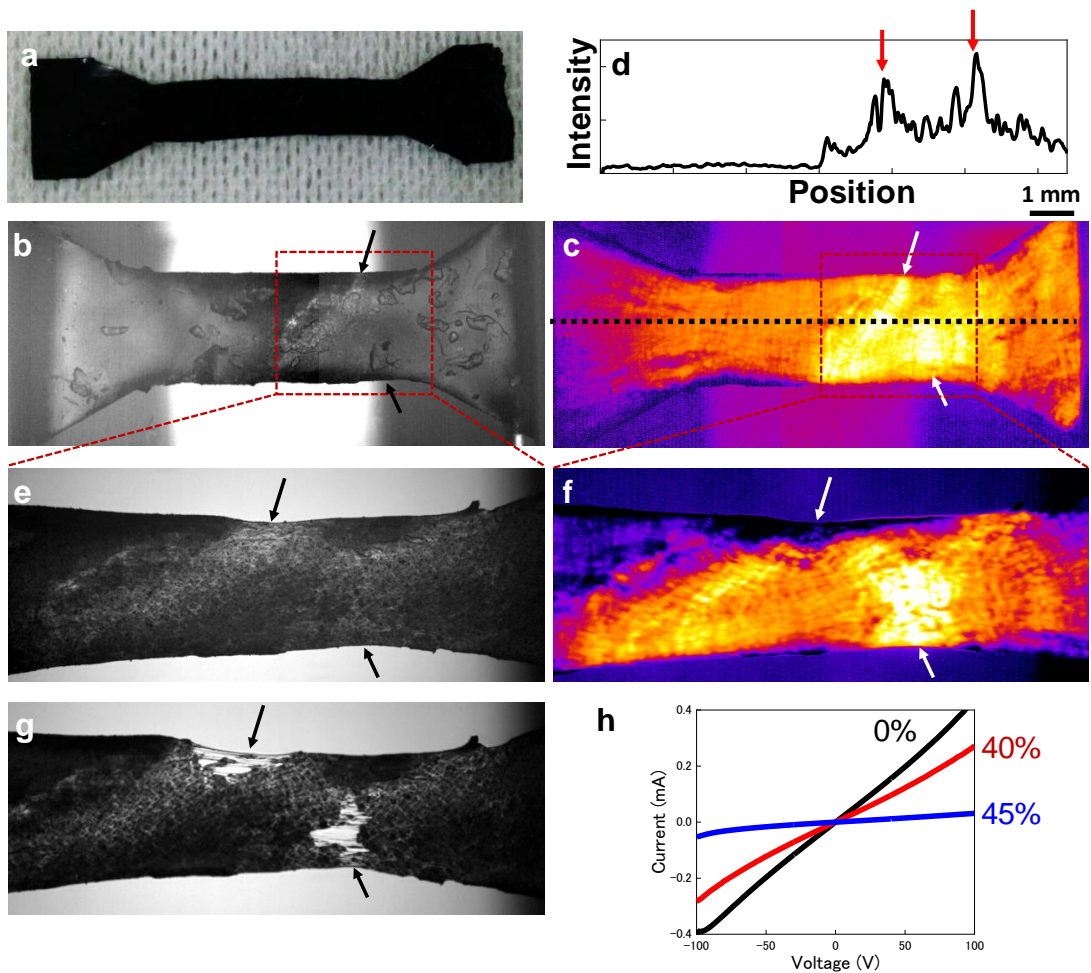

**Figure. 5** (**a**) Optical microscope image of sample made from same rubber sheet measured in tensile test. (**b-g**) The IR and LIT intensity images obtained during tensile testing at initial condition (b,c) and at 40% strain (e,f). (**d**) Cross section of intensity profile at initial state shown in (c) (dashed line). The arrows indicate the strong heating points corresponding to the high-resistance region. (**g**) The sample has broken at 50% strain. All arrow corresponds to the high-intensity region at initial condition. (**h**) The I-V characteristics in each tensile condition. The two-terminal resistance was changed from 25.9 kΩ to 1.9 MΩ.

**Table of Contents Graphic**

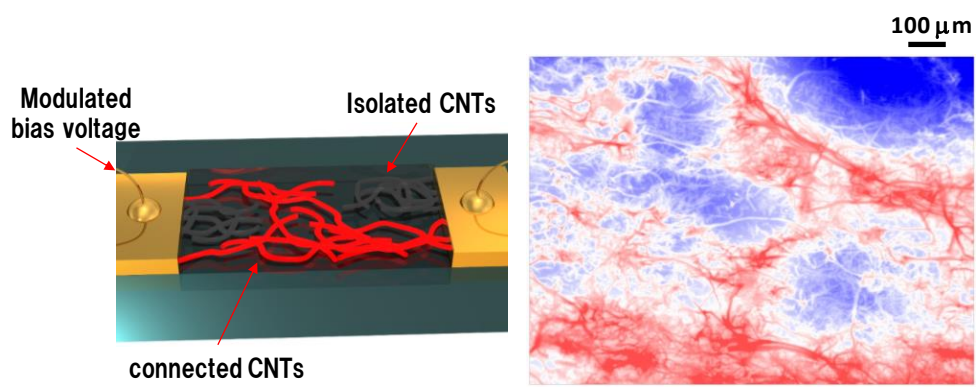

**ACKNOWLEDGEMENTS**

**Funding:** This paper is based on results obtained in a project commissioned by the New Energy and Industrial Technology Development Organization (NEDO).

**Author contributions:** T.M performed sample fabrication and measurements using the LIT method. S.A and T.Y fabricated composite materials. All authors discussed the results and wrote the manuscript.

# Supplementary information

## S-1. The fundamental process of LIT method

In this manuscript, we have applied the real-time lock-in thermography (LIT) technique to measure the current density distribution and randomly networked structures via the Joule heating from the sample surface. This method is basically similar to the "dual phase lock-in technique" in conventional electric or optical measurement systems. In the conventional systems, the lock-in process is applied for improving the signal noise ratio (S/N). In our case, the LIT method is mainly used as frequency space separations of various heat components.

In this time, we used the THERMOS-1000 (Hamamatsu Photonics) as base setup of the lock-in thermography (LIT) systems. The IR camera is consisted with an InSb detector cooled by a Stirling cooler. The maximum pixel size is (640 x 512) at 25 Hz measurement. The IV characteristic and ac bias voltage are applied by the semiconductor parameter analyzer (B1500, Keysight Technologies). The detected IR wavelength was ranged from 3.0 to 5.0 μm. This range is suitable at room temperature measurements from the estimation based on the Stefan-Boltzmann's law.

The Joule heating is generated by applied bias voltages with rectangular 25 Hz repeated signals. The original data is obtained as IR images in each frame. These IR images provide the thermal intensity depending on the ac bias voltages. In the real time LIT systems, the lock-in calculation is done in each cycle step. After calculation in each step, the original IR frame data is removed from the PC memories for avoiding the huge data creations for long time accumulations. By this system, we can select the suitable accumulation time beyond limitation of the storage and memory capacity of PC systems. Then, finally, we obtain the intensity and phase images constructed by calculated all pixel data.

Figure S1 shows the schematic picture of the calculation process in this method. The intensity R having phase $\theta$ is estimated complex vector 0° and 90° as shown in Fig. S1. From this process, the intensity R is always calculated without phase matching process. This is similar to the dual phase lock-in technique in conventional method. Actually, the heat cycle line-shape is lost by this process. The averaged phase value is also calculated, and this is related to the delay of heating cycles.

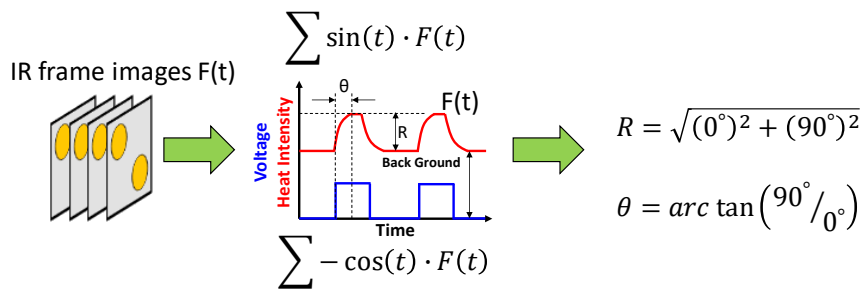

**Figure S1 | Schematic picture of the LIT method process**

The schematic picture of the LIT method. The raw data is observed as IR images with IR camera systems. The intensity and phase parameters are corresponding the vector length and angle in the complex coordinate space.

## S2. The phase separation images in the thin sample.

The difference of heating phase is clearly visualized by the mixing of intensity and phase in each pixel. Figure. S2 shows the results of different timing image calculated as $I_{LIT} = R \cos \theta$. In the low frequency ac bias conditions, the ac current is immediately excited and generate the Joule heating at the current path. On the other hand, the non-active region without excited currents has delay heat component due to the heat conduction from the heat source. Therefore, the faster on-phase component indicated the real current density distributions. The disappeared region in the on-phase image are observed in delay component images. These regions have heat component having same frequency of the ac bias voltages. This expectation is confirmed by the close-up images as shown in Fig. S2b - 2d. In this image, the broken CNT is observed, and this CNT's signal is completely suppressed in the delay component image. This is naturally understood that the current path should be interrupted at the broken point. In these non-active region, the LIT intensity contained only the thermal conductive component without a Joule heating source. Therefore, we compared the LIT intensity and theoretical estimated curves in these regions.

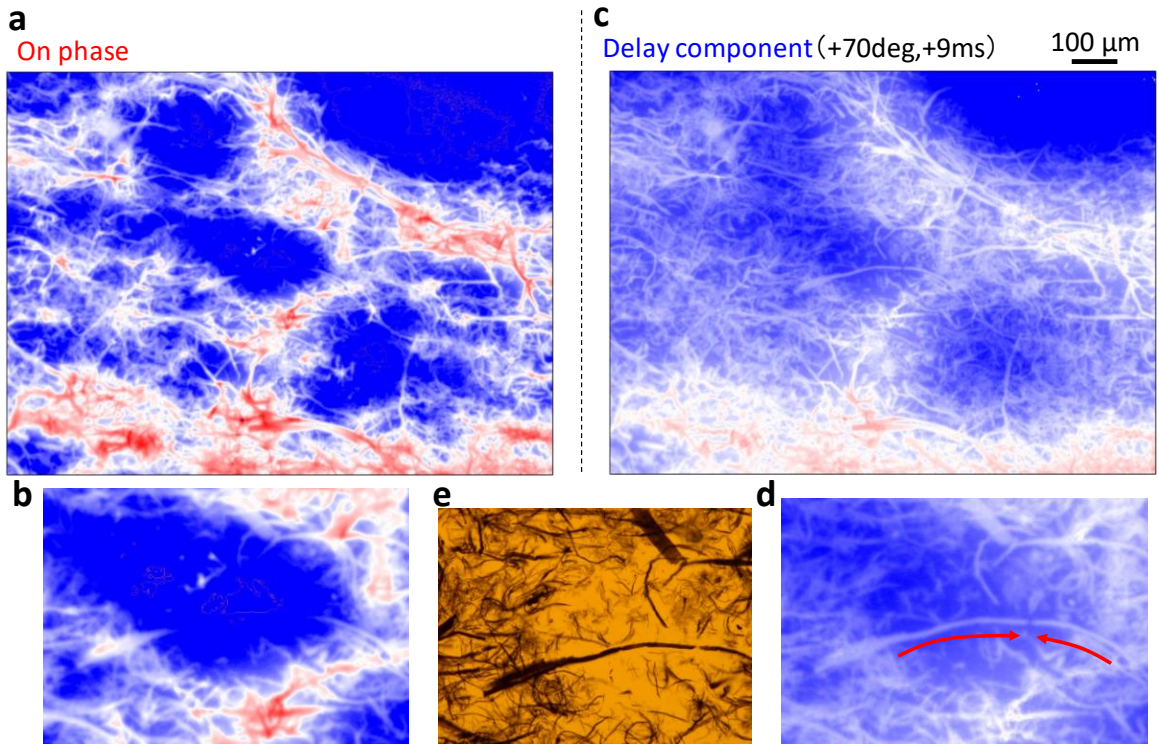

**Figure S2 | The phase separation images in the thermally thin sample**

The phase separation LIT images in higher magnification measurement. **a** and **b**, The on-phase component images having same frequency and phase with an ac bias voltages. **c** and **d**, The delay component images. In these images, the thermal conductive signals were observed existed around the Joule heating source area. **e**, The close-up image by optical microscope at same position of **b** and **d**.

**S3. The theoretical estimation based on the heat conduction in thermally thin samples.**

In the LIT method, the temperature distributions are observed via the heat radiation at the sample surface. These temperature differences are originated from the Joule heating and thermal conduction to the environment from the heat source. Therefore, the intensity of LIT images should be related to the thermal conduction in microscopic pictures. Especially, in the ac bias conditions, the heat and temperature also have oscillating time dependent components.

The temperature dumping of heat conduction with oscillation heat source is represented as:

$$\Lambda = \sqrt{\frac{2\alpha}{f}}$$

here, $\Lambda$ is the thermal diffusive length, $\alpha$ is the thermal diffusivity and $f$ is the frequency. The intensity of conductive temperature is decreased as $1/e$ at this thermal diffusive length. For estimation of the LIT intensity, the sample thickness is one important factor for thermally thickness. In the case of thermally thin sample (sample thickness is thinner than the $\Lambda$), the thermal conduction should be occurred including multiple reflections at the interfaces. These multiple reflections are considered as the mirror heat source. Therefore, the point like heat source should act as cylindrical-shape heat sources. In these description, the thermal conduction should be described based on the one kind of the Bessel functions.

$$T(r,t) = AK_0\left(r\sqrt{\frac{iC_P n\omega}{\lambda}}\right)e^{i\omega t} = A\left(ker\left(\frac{r\sqrt{2}}{\Lambda}\right) + ikei\left(\frac{r\sqrt{2}}{\Lambda}\right)\right)e^{i\omega t}$$

here is, $C_P$ is specific heat, $\lambda$ is thermal conductivity, $n$ is material density and $r$ is distance from the heat source. Figure S4 shows the estimated curves of various typical materials. The all curve shows strongly intensity dumping near the heat source. This strongly dumping curve is important for higher resolutions of the LIT method. This result suggests that the LIT method has high spatial resolution between a heat source and non-active regions.

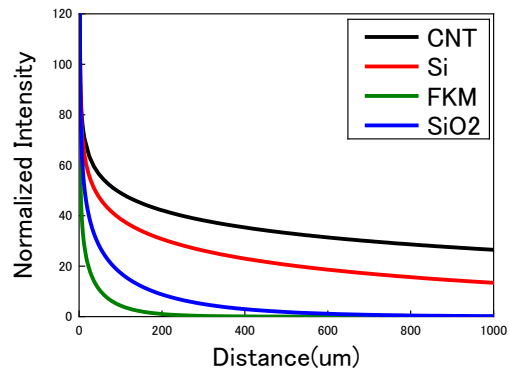

**Figure S3 | The estimated theoretical curves of various typical materials.**

The thermal dumping curves calculated by the thermal conduction equation. Each carve is corresponding to the typical material, single CNT (black), silicon (red), silicon di-oxide (blue) and FKM rubber (green). The CNT bundle had similar thermal conductivity value with the silicon.

### S4. The quantitative analysis of LIT intensity distributions

The LIT intensity distributions should be depended to the homogeneity of CNT network structures. By increasing the CNT concentrations, the uniform network structures are easily realized based on the percolation models. As shown in Fig. S4, the lower density sample indicated strongly asymmetric line-shape. Especially, the long tail line-shape is caused by the bottleneck structures as shown in Fig. 2. In this case, it is difficult to fit the curve by the normal distribution function as shown in Fig. S4**a**. On the other hand, the higher density sample shows almost completely normal distributions. The full-width and half maximum (FWHM) is also smaller than the lower density samples.

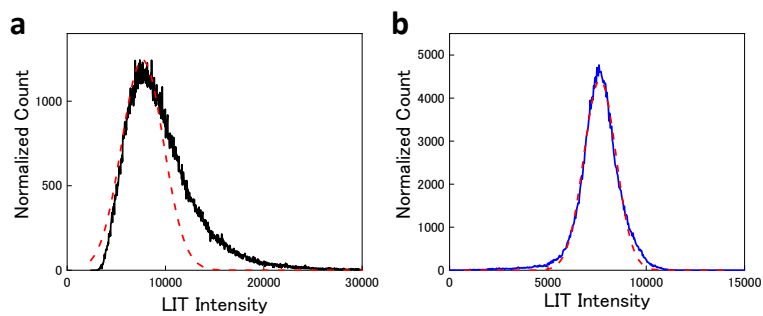

**Figure S4 | The LIT intensity distribution and normalized distribution fitting curves.**

**a** and **b**, The LIT intensity distributions are shown with the normalized distribution fitting curves. The higher density sample (1.0 wt%) had almost perfect fitting with normal distributions as shown in **b**. The lower density sample (0.25 wt%) had asymmetric distributions due to additional higher intensity components.

**S5. The current flow direction dependence in same network structures.**

In the case of the randomly connected network structures, the current path should be chosen for minimize the total impedance between electrodes. These electric circuits are consisted with the nano-material's resistance and junctions between nano-materials. Therefore, it is possible that the local low resistance region is not effective in different bias configurations.

Figure. S5 shows the current flow direction dependence observed in same sample and measurement area. In the cross-shape sample, the non-active regions are clearly observed depending on the current flow directions as shown in Fig. S5a and 5c. These results showed that the non-active region does not contribute to reduce the total impedance by their long current path. This simple rule is also effective in higher magnification measurement for a few µm scale images. For example, in the black circles potion, the strong LIT intensity is observed in horizontal current conditions. In the case of vertical current flow conditions, however, the current path is completely different and avoided in this region. This difference should be originated from the difference of orientation and connectivity of CNTs. These results showed that the LIT method have possibility to analyze the detail of transport mechanism in these percolated network structures.

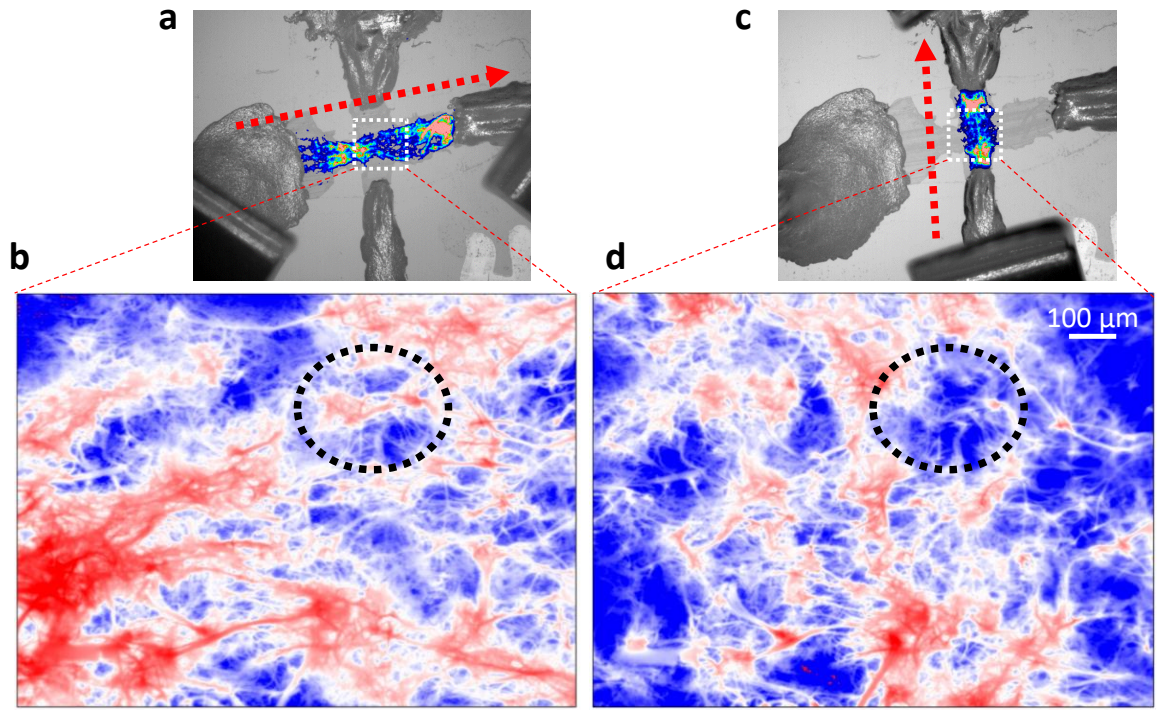

**S5| The current flow direction dependent LIT images on same sample and area.**

**a** and **c**, The overlay images of LIT and IR image observed by different current flow direction. **b** and **d**, The high magnification LIT images in each current flow direction. The black dot circles indicate the same potion.